\documentclass[iop,apj]{emulateapj}
\usepackage{apjfonts}

\begin{document}

\epsscale{1.1}

\title{The Quenching of the Ultra-Faint Dwarf Galaxies in the Reionization 
Era\altaffilmark{1}}
\shorttitle{Quenching of the Ultra-Faint Dwarf Galaxies}

\author{
Thomas M. Brown\altaffilmark{2}, 
Jason Tumlinson\altaffilmark{2}, 
Marla Geha\altaffilmark{3}, 
Joshua D. Simon\altaffilmark{4}, 
Luis C. Vargas\altaffilmark{3}, 
Don A. VandenBerg\altaffilmark{5}, 
Evan N. Kirby\altaffilmark{6}, 
Jason S. Kalirai\altaffilmark{2,7}, 
Roberto J. Avila\altaffilmark{2}, 
Mario Gennaro\altaffilmark{2}, 
Henry C. Ferguson\altaffilmark{2} 
Ricardo R. Mu\~noz\altaffilmark{8}, 
Puragra Guhathakurta\altaffilmark{9}, 
and
Alvio Renzini\altaffilmark{10} 
}

\altaffiltext{1}{Based on observations made with the NASA/ESA {\it Hubble
Space Telescope}, obtained at the Space Telescope Science Institute, which
is operated by the Association of Universities for Research in Astronomy, 
Inc., under NASA contract NAS 5-26555.  These observations are associated
with program GO-12549.}

\altaffiltext{2}{Space Telescope Science Institute, 3700 San Martin Drive,
Baltimore, MD 21218, USA;  
tbrown@stsci.edu, tumlinson@stsci.edu, jkalirai@stsci.edu, avila@stsci.edu,
ferguson@stsci.edu, gennaro@stsci.edu}

\altaffiltext{3}{Astronomy Department, Yale University, New Haven, CT
06520, USA; marla.geha@yale.edu, luis.vargas@yale.edu}

\altaffiltext{4}{Observatories of the Carnegie Institution of Washington, 
813 Santa Barbara Street, Pasadena, CA 91101, 
USA; jsimon@obs.carnegiescience.edu}

\altaffiltext{5}{Department of Physics and Astronomy, 
University of Victoria, P.O. Box 1700, STN CSC, Victoria, BC, V8W 2Y2, Canada; 
vandenbe@uvic.ca}

\altaffiltext{6}{California Institute of Technology, 1200 East
California Boulevard, MC 249-17, Pasadena, CA 91125, USA;
enkirby@gmail.com}

\altaffiltext{7}{Center for Astrophysical Sciences, Johns Hopkins
University, Baltimore, MD, 21218}

\altaffiltext{8}{Departamento de Astronom\'ia, Universidad de Chile, 
Casilla 36-D, Santiago, Chile; rmunoz@das.uchile.cl}

\altaffiltext{9}{UCO/Lick Observatory and Department of Astronomy and 
Astrophysics, University of California, Santa Cruz, CA 95064, USA; 
raja@ucolick.org}

\altaffiltext{10}{Osservatorio Astronomico, Vicolo Dell'Osservatorio 5, 
I-35122 Padova, Italy; alvio.renzini@oapd.inaf.it}

\submitted{Accepted for publication in The Astrophysical Journal}

\begin{abstract}

We present new constraints on the star formation histories of six
ultra-faint dwarf galaxies: Bootes~I, Canes Venatici~II, Coma
Berenices, Hercules, Leo~IV, and Ursa Major~I.  Our analysis employs a
combination of high-precision photometry obtained with the Advanced
Camera for Surveys on the {\it Hubble Space Telescope},
medium-resolution spectroscopy obtained with the DEep Imaging
Multi-Object Spectrograph on the {\it W.M. Keck Observatory}, and
updated {\it Victoria-Regina} isochrones tailored to the abundance
patterns appropriate for these galaxies.  The data for five of these
Milky Way satellites are best fit by a star formation history where at least 75\%
of the stars formed by $z \sim 10$ (13.3~Gyr ago).  All of the galaxies
are consistent with 80\% of the stars forming by $z \sim 6$ (12.8~Gyr
ago) and 100\% of the stars forming by $z \sim 3$ (11.6~Gyr ago). The
similarly ancient populations of these galaxies support the hypothesis
that star formation in the smallest dark matter sub-halos was
suppressed by a global outside influence, such as the reionization of
the universe.

\end{abstract}


\section{Introduction}

One of the primary quests of astronomy is understanding the formation
of structure in the universe.  In this regard, the $\Lambda$ Cold Dark
Matter ($\Lambda$CDM) cosmological model is consistent with many
observable phenomena, but there are discrepancies at small scales
(Kauffmann et al.\ 1993).  Specifically, $\Lambda$CDM predicts many
more dark-matter sub-halos than the number observed as dwarf galaxies
(e.g., Moore et al.\ 1999; Klypin et al.\ 1999) -- the 
``missing satellite'' problem.  As
one way of rectifying this problem, Bullock et al.\ (2001) put forth
the idea that reionization could have suppressed star formation in the
smallest DM sub-halos (see also Babul \& Rees 1992), essentially by
boiling the gas out of their shallow potential wells.  The dearth of
stars would then make these sub-halos difficult or impossible to detect.  
Building upon
this hypothesis, Ricotti \& Gnedin (2005) proposed that dwarf galaxies
could follow one of three evolutionary paths: ``true fossils'' that
formed most of their stars prior to reionization, ``polluted fossils''
with star formation continuing beyond reionization, and ``survivors''
that largely formed their stars after reionization.  It is now common
for galaxy formation models to alleviate the missing satellite problem
by truncating the star formation in DM halos below some nominal
mass threshold, sometimes termed the ``filtering mass,'' with this threshold 
tuned to match the observations (e.g., Tumlinson 2010;
Mu$\tilde{\rm n}$oz et al.\ 2009; Bovill et al.\ 2009, 2011a, 2011b;
Koposov et al.\ 2009; Li et al.\ 2010; Salvadori et al.\ 2009, 2014).

\begin{table*}[t]
\begin{center}
\caption{{\it HST} ACS Observations}
\begin{tabular}{lcccccrrr}
\tableline
                  &         &         &                &               & Field\tablenotemark{c}      & \multicolumn{2}{c}{Exposure per tile}  &  \\
                  & R.A.\tablenotemark{a}    & Dec.\tablenotemark{a}    & $(m-M)_V$\tablenotemark{b}      & E$(B-V)$\tablenotemark{b}      & Contamination   & F606W      & F814W                   &  \\
Name              & (J2000) & (J2000) & (mag)          & (mag)         & (\%)            &(s)       & (s)                       & Tiles \\
\tableline
Bootes I          & 14:00:04&+14:30:47& 19.11$\pm$0.07 & 0.04$\pm$0.01 & 7.9             & 2,340  &  2,200 &  5\\
Canes Venatici II & 12:57:10&+34:19:23& 21.04$\pm$0.06 & 0.04$\pm$0.01 & 2.2             & 20,850 & 20,850 &  1\\
Coma Berenices    & 12:27:21&+23:53:13& 18.08$\pm$0.10 & 0.04$\pm$0.01 & 24              & 2,340  &  2,200 & 12\\
Hercules          & 16:31:05&+12:47:07& 20.92$\pm$0.05 & 0.09$\pm$0.01 & 6.1             & 12,880 & 12,745 &  2\\
Leo IV            & 11:32:57&-00:31:00& 21.12$\pm$0.07 & 0.08$\pm$0.01 & 3.7             & 20,530 & 20,530 &  1\\
Ursa Major I      & 10:35:04&+51:56:51& 20.10$\pm$0.05 & 0.05$\pm$0.01 & 17              & 4,215  &  3,725 &  9\\
\tableline

\multicolumn{9}{l}{$^{\rm a}$Center of ACS observations.} \\ 
\multicolumn{9}{l}{$^{\rm b}$Apparent distance moduli and extinctions are determined from fits to the ACS data.} \\
\multicolumn{9}{l}{$^{\rm c}$Contamination near the upper MS, based upon the Besan\c{c}on Galaxy model (Robin et al.\ 2003).} \\
\end{tabular}
\end{center}
\end{table*}

Over the same time period, wide-field surveys revealed the existence
of additional dwarf satellites around the Milky Way (e.g., Willman et
al.\ 2005; Zucker et al.\ 2006; Belokurov et al.\ 2007) and Andromeda
(e.g., Zucker et al.\ 2004, 2007; McConnachie et al.\ 2009; Majewski
et al.\ 2007; Irwin et al.\ 2008; Martin et al.\ 2009).  The ultra-faint
dwarf (UFD) galaxies have luminosities of $M_V > -8$~mag ($M_* \lesssim
10^4~M_\odot$; Martin et al.\ 2008b), and thus most
are fainter than the typical globular cluster.  Photometric and
spectroscopic observations of the UFD galaxies
have shown that they are excellent candidates for demonstrating the
existence of fossil galaxies.  Color-magnitude diagrams (CMDs)
indicate the UFDs are generally dominated by old ($>$10~Gyr)
populations (e.g., Sand et al.\ 2009, 2010; Okamoto et al.\ 2008,
2010, 2012; de Jong et al.\ 2008b; Hughes et al.\ 2008; Martin et
al.\ 2008a; Greco et al.\ 2008; Mu\~noz et al.\ 2010; Weisz et
al.\ 2014b), while spectroscopy of their giant stars indicates low
metallicities, but with a dispersion significantly larger than the
measurement errors (Frebel et al.\ 2010; Norris et al.\ 2010; Kirby et
al.\ 2008, 2011, 2013).  The internal kinematics from such
spectroscopy also imply large mass-to-light ratios ($M/L_V \gtrsim
100$; e.g., Kleyna et al.\ 2005; Mu\~noz et al.\ 2006; Martin et
al.\ 2007; Simon \& Geha 2007).  Because even the most massive
globular clusters have $M/L_V$ ratios consistent with little to no
dark matter (e.g., Baumgardt et al.\ 2009; van de Ven et al.\ 2006;
Bradford et al.\ 2012),
the high $M/L_V$ in the UFDs is one of the characteristics that marks
them as galaxies, instead of star clusters, despite their low
luminosities.  Another distinction with most star clusters is the fact
that the stellar populations of galaxies exhibit spreads in age and
metallicity.  Given their low metallicities, old ages, faint luminosities, and
high $M/L_V$ ratios, the UFDs are an excellent laboratory to search
for reionization signatures in the star formation history (SFH) of
small DM sub-halos, and to assess the possible solutions to the
missing satellite problem.

In this paper, we present new constraints on the SFHs of six UFD
galaxies: Bootes I (Boo~I), Canes Venatici~II (CVn~II), Coma Berenices
(Com~Ber), Hercules, Leo~IV, and Ursa Major~I (UMa~I). Our analysis
focuses on high-precision photometry, from the Advanced Camera for
Surveys (ACS) on the {\it Hubble Space Telescope (HST)}, and new
medium-resolution spectroscopy, from the DEep Imaging Multi-Object
Spectrograph (DEIMOS) on the {\it W.M. Keck Observatory}.  We
interpret these data using a new isochrone grid generated with the
{\it Victoria-Regina} code (VandenBerg et al.\ 2012), employing the
latest physics, and assuming abundance profiles appropriate to the
extremely metal-poor populations of the UFDs.

\section{Observations and Data Reduction}

\subsection{Hubble}

\subsubsection{Observations}

From Aug 2011 through Jun 2012, we obtained deep optical images of
each galaxy in our sample (Table~1) using the F606W and F814W filters
on ACS (GO-12549; PI Brown).  A preliminary analysis of the earliest observations in this
program was given by Brown et al.\ (2012).
These galaxies were chosen to provide a representative sample
of UFDs with integrated luminosities well below those of the classical
dwarf spheroidals, but bright enough to provide sufficient numbers of
stars for the SFH analysis.  Specifically, the goal was to obtain
photometry with a high signal-to-noise ratio ($SNR \sim 100$) for
$\gtrsim$100 stars within 1~mag of the main sequence (MS) turnoff,
thus cleanly defining the upper MS, subgiant branch (SGB), and lower
red giant branch (RGB), and allowing sub-Gyr precision in 
relative ages.  The turnoff has long been a reliable clock
for the dating of stellar populations (e.g., Iben \& Renzini 1984;
VandenBerg et al.\ 1990), becoming fainter and redder at increasing
age, but the changes are subtle at old ages.  For example, at
[Fe/H]=$-2.4$ and 12~Gyr, an age increase of 1~Gyr shifts the turnoff
0.09~mag fainter in $m_{\rm 814}$ and 0.01~mag redder in $m_{\rm 606}
- m_{\rm 814}$.  Although there is no age information below the
turnoff, obtaining high SNR photometry at the turnoff produces
photometry with a faint limit below 0.5~$M_\odot$ on the MS, enabling
measurements of the stellar initial mass function (IMF; see Geha et
al.\ 2013).  Because their distances and apparent sizes span a wide
range, the observing strategy for each galaxy was tailored to obtain
photometry of similar quality in each galaxy, surveying a wide but
shallow area in the relatively nearby satellites (e.g., Com~Ber), and
a narrow but deep pencil beam in the more distant satellites (e.g., CVn~II).

\subsubsection{Reduction}

The images were processed with the latest pipeline updates, including
a pixel-based correction (version 3.2) for charge-transfer inefficiency (CTI; 
Anderson \& Bedin 2010) resulting from radiation damage to the ACS
detectors.  The individual exposures were dithered to enable
resampling of the point spread function (PSF), mitigation of
detector artifacts (hot pixels, dead pixels), and cosmic ray rejection.  The
exposures for each tile in each band were coadded with the {\sc
  drizzle} package (Fruchter \& Hook 2002), using the {\sc tweakshifts}
routine to iteratively solve for the offsets between individual images.
This process produced
geometrically-correct images with a scale of 0.035$^{\prime\prime}$
pixel$^{-1}$ and an area of approximately $210^{\prime\prime} \times
220^{\prime\prime}$.

\subsubsection{Photometry}

We performed both aperture and PSF-fitting photometry using the
DAOPHOT-II package (Stetson 1987), assuming a spatially-variable PSF
constructed from isolated stars.  The final catalog combined aperture
photometry for stars with photometric errors $<$0.01~mag and
PSF-fitting photometry for the rest, with both normalized to an infinite
aperture.  Due to the scarcity of bright stars, the uncertainty in the
normalization to an infinite aperture is $\sim$0.02~mag.  For the
three nearest galaxies (Com~Ber, Boo~I, and UMa~I), the scarcity of
bright stars in any individual tile hampered the construction of an
accurate PSF model, so a spatially-dependent PSF model for each galaxy
was constructed from all of the tiles in a given band, selecting isolated
bright stars from each tile.  For Hercules,
there were enough stars to construct an independent PSF model in each
of the two tiles, but then the normalizations of those PSF models were
adjusted to give agreement between the two tiles.  Similarly, the
single tiles obtained in Leo~IV and CVn~II were sufficiently populated
to construct spatially-dependent PSF models for each. Our photometry
is in the STMAG system: $m= -2.5 \times $~log$_{10} f_\lambda -21.1$.
The catalogs were cleaned of background galaxies and stars with poor
photometry using the $\chi^2$ of the PSF fitting, the PSF sharpness,
and photometric errors.  Stars were also rejected if they fell within
the wings of brighter neighbors or within the extent of a background
galaxy.  After all the cuts were applied, between 12\% and 35\% of the
sources were rejected from each catalog, largely near the faint limit.

Transformation from the {\it HST} photometric system to a ground-based
system incurs significant systematic errors, as explored by Sirianni
et al.\ (2005).  For this reason, a direct comparison between our
photometry and previously-published catalogs is of limited utility.
However, for one galaxy in our sample (CVn~II), a catalog with bands
that overlap with our own ($V$ and $I$) is publicly available (Sand et
al.\ 2012).  The transformations in Sirianni et al.\ (2005) do not
reflect the updates to the ACS calibration after the last {\it HST}
servicing mission, but we can derive our own transformations, using
the available throughput curves in each system and the synthetic
spectral library of Gustafsson et al.\ (2008).  Doing so, we find that
the photometry of the brightest stars in our catalog (20--23~mag)
agrees with the Sand et al.\ (2012) photometry of these same stars at
the level of 0.03~mag.  This comparison demonstrates that there are no
gross calibration differences between the {\it HST} photometry and
previously published photometry from the ground.

\begin{figure*}[t]
\plotone{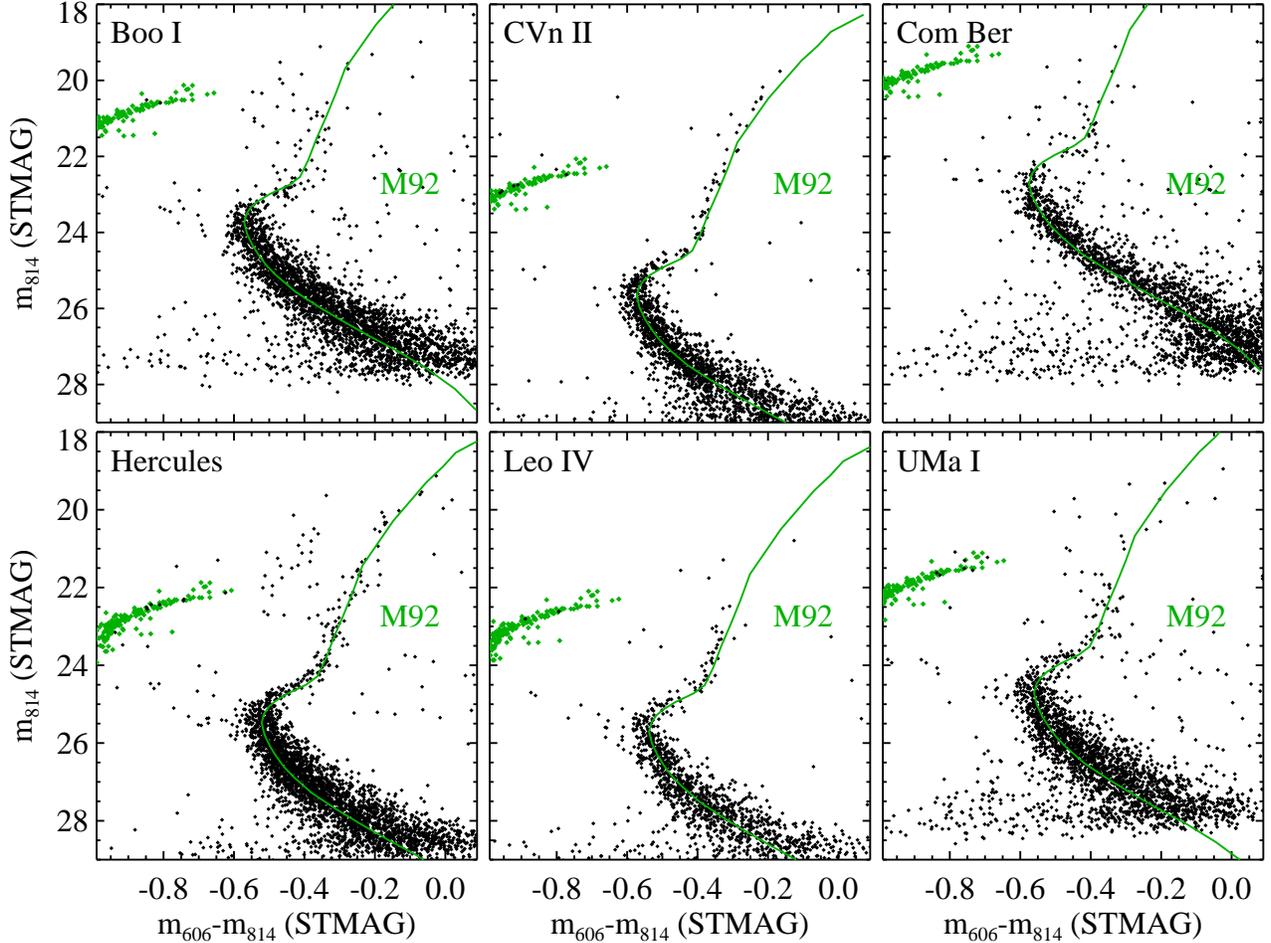}
\caption{The CMD of each UFD in our sample (black points).
  For reference, we show
  the empirical ridge line for the MS, SGB, and RGB in M92 (green
  curve), along with the HB locus in M92 (green points).  The M92
  fiducial has been placed at the distance and reddening for
  each galaxy (Table~1), matching the luminosity of HB stars
  and the color of the lower MS stars.  Because the CMD of each galaxy
  looks, to first order, like that of a ancient metal-poor globular
  cluster, the stellar population of each galaxy is dominated by
  ancient metal-poor stars.  The CMDs of these galaxies are all extremely
  similar to one another, implying they have similar stellar populations
  and star formation histories.}
\end{figure*}

To properly account for the photometric errors and completeness, we
performed artificial star tests using the same photometric routines
that were employed for the photometric catalogs.  During such tests,
one does not want to affect the crowding of the images, so small
numbers of artificial stars were repeatedly added to each image and
then blindly recovered, until there were over 5,000,000 artificial
stars for each galaxy.  To ensure that the noise in the artificial
stars accurately represented that in the data in this high SNR regime,
we included detector effects that would not be experienced by an
artificial star simply inserted into the images and recovered.  We
assumed a residual flat fielding error of 1\% (Gonzaga et al.\ 2014),
and inserted artificial stars with the reduction in signal appropriate
for the CTI that a real star would encounter at that signal level and
background in each image (using the forward-modeling CTI software that
is included in the CTI correction package).  Although CTI losses in
both real and artificial stars can be corrected to the appropriate
flux level, these corrections do not recover the loss of SNR, because
measurements still have the shot noise on the reduced signal.
Neglecting this effect in the artificial star tests would make the
photometry of artificial stars slightly less noisy than that of the
real stars at the same magnitude.

\subsubsection{Color-Magnitude Diagrams}

The CMD of each galaxy in our survey is shown in Figure~1. Two aspects
of these CMDs are immediately apparent.  First, the tight stellar
locus of each CMD resembles that of a Galactic globular cluster, as
will be discussed in the next section.  Second, the CMDs all appear
extremely similar to each other, implying the population ages and
metallicities are also similar.  In Figure~2, we show the composite
CMD for all 6 galaxies in our sample, each shifted to the same
distance and reddening (see \S3.1), and focused on the CMD region most
sensitive to age (i.e., the MS turnoff and SGB).  To the eye, each UFD
appears to be dominated by an ancient metal-poor population.

\begin{figure}[t]
\plotone{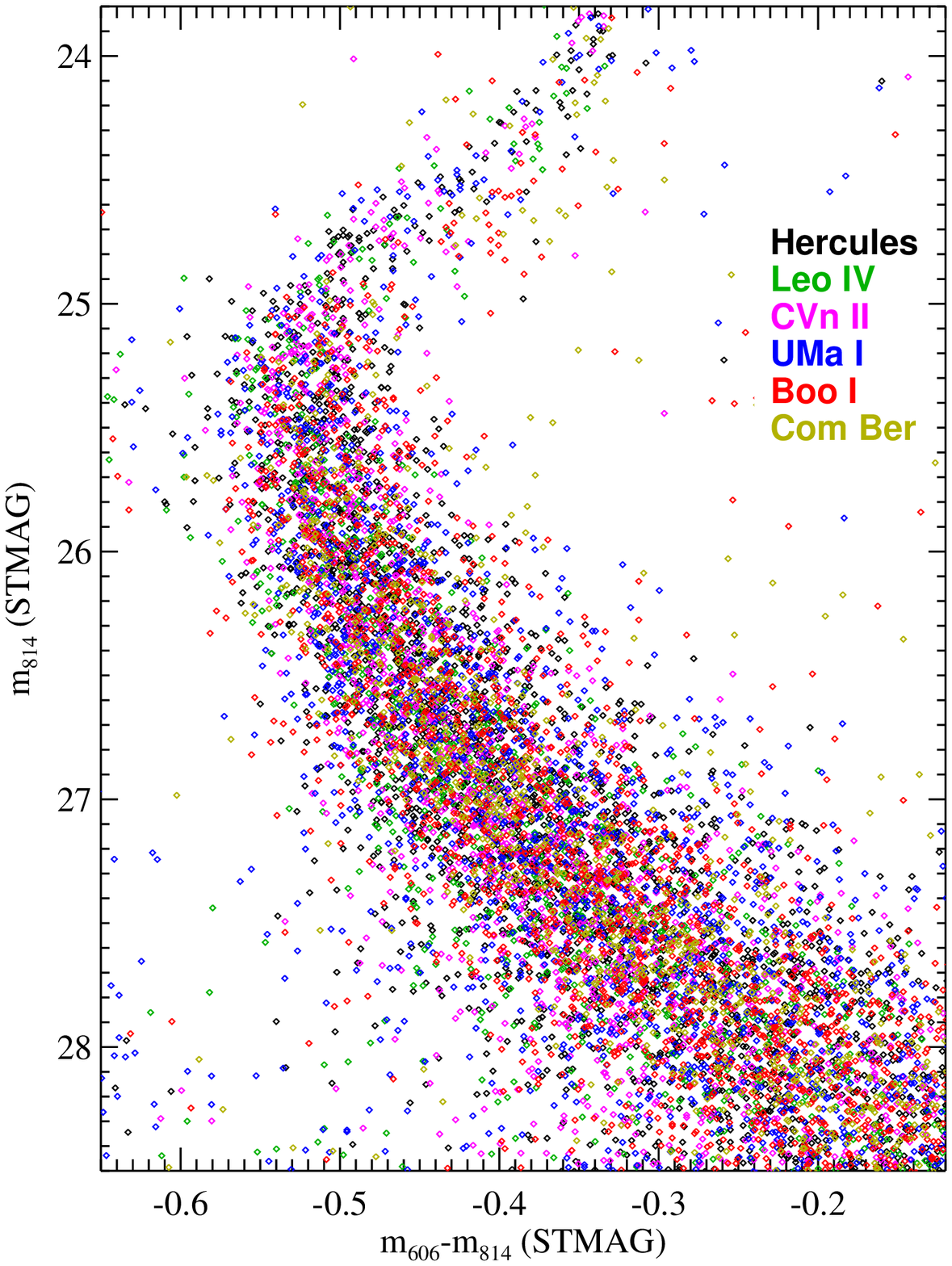}
\caption{The CMD of each UFD (colored points), shifted
  to the distance and reddening of Hercules, and zoomed
  into the CMD region most sensitive to age.
  The similarities of the 6 CMDs imply that the UFD populations 
  are extremely similar in age and metallicity.}
\end{figure}

Further inspection of the CMDs reveals other details worth noting.
Field contamination is high for those relatively nearby galaxies that
were observed in several tiles (Com~Ber, Boo~I, and UMa~I), and is
apparent from the scattering of stars beyond the main stellar locus.
The level of field contamination in the vicinity of the upper MS,
where we fit the SFH,
can be estimated by transforming the
Besan\c{c}on Galaxy model (Robin et al.\ 2003) to the ACS bands used
here, and is reported in Table~1.  The contamination depends upon
the surface brightness of each galaxy, and scales with
the number of tiles observed. There are also a few blue straggler
(BS) stars apparent in each CMD, falling to the blue and extending
brighter than the dominant MS turnoff.  Although BS stars are common
in ancient populations, they can mimic a much younger sub-population.
For example, the turnoff mass at 12--13~Gyr is $\sim$0.8~$M_\odot$, but 
BS stars can be up to twice as large, which would not normally appear on
the MS for populations older than 2~Gyr.
The BS frequency is generally expressed relative
to that of horizontal branch (HB) stars.  In globular clusters,
$N_{BS} / N_{HB}$ typically ranges from 0.1 to 1 (e.g., Piotto et
al.\ 2004; Ferraro et al.\ 2014).  In the Galactic halo, Preston \& Sneden (2000) find a
much higher ratio: $N_{BS} / N_{HB}$~=~4.4.  In low-luminosity dwarf
galaxies, Momany et al.\ (2007) find $N_{BS} / N_{HB}$ ranging from 1
to 4, and Santana et al.\ (2013) found the frequency of BS stars to be
similar in UFDs and the classical dwarf spheroidals.
For the UFD CMDs here, our statistics on both BS stars and HB
stars are too poor to give strong constraints on the BS frequency;
assuming that HB stars cannot fall more than 0.2~mag below the 
expected HB locus, we estimate that
$N_{BS} / N_{HB} \sim 2$.  BS stars are largely excluded from our fits, except
for any that might lie immediately adjacent to the dominant MS.

\subsection{Keck}

\subsubsection{Observations}

Metallicities for limited samples of stars in five of the six UFDs
targeted with {\it HST} were determined by Kirby et al.\ (2008),
Kirby et al.\ (2011), and Vargas et al.\ (2013), based on the
medium-resolution (1.37~\AA\ FWHM)
{\it Keck} spectroscopy of Simon \& Geha (2007).  However, fewer than 16
measurements were available in every galaxy except UMa~I, and Simon \&
Geha (2007) did not observe Boo~I at all.  To improve the constraints
on the metallicity distributions (which, in turn, improve the
constraints on the ages determined from the {\it HST} photometry), we
obtained new Keck/DEIMOS spectroscopy for larger samples of stars in
Leo~IV, Com~Ber, CVn~II, Boo~I, and Hercules.  On the nights of 2013
March 10--11, 2013 April 12, and 2013 May 3--4, we observed a total of
13 slit masks, with typical integration times of 1--3~hr.  Conditions
during the observations ranged from good to poor.  Mask design and
calibration procedures followed those established by Simon \& Geha
(2007), Geha et al.\ (2009), and Simon et al.\ (2011).  For Hercules
and Com~Ber, we also include several slit masks observed in 2010 and
2011 that have not yet been published.  Note that the spectroscopic
samples were targeted and analyzed using ground-based photometry,
instead of the {\it HST}/ACS photometry described above.
The small {\it HST} field of view is insufficient for multi-object 
spectroscopic selection, and much of the Keck analysis preceded 
the {\it HST}/ACS observations.

\subsubsection{Reduction}

The spectroscopic data were reduced with our slightly modified versions of the
DEEP2 pipeline (Cooper et al.\ 2012), as described in Simon \& Geha (2007).
We measure stellar metallicities using the large
number of neutral iron lines included in our spectral range ($6300 <
\lambda < 9100$~\AA). We match each spectrum against a grid of
synthetic spectra sampling a wide range in [Fe/H], [$\alpha$/Fe],
log~$g$, and $T_{\rm eff}$ (Kirby 2011).  Prior to
fitting, we degrade the synthetic spectra to the DEIMOS resolution.  We
excise wavelength regions affected by telluric contamination, strong
sky emission lines, and regions improperly synthesized due to NLTE
effects (Ca triplet and \ion{Mg}{1} $\lambda$8807).

\subsubsection{Metallicities}

We determine the best-fitting $T_{\rm eff}$ and [Fe/H] values
simultaneously from $\chi^2$ minimization of the pixel-by-pixel flux
difference between the observed spectra and the synthetic grid, using
only spectral regions sensitive to variations in Fe abundance. We
separately fit [$\alpha$/Fe] using regions sensitive to Mg, Si, Ca,
and Ti variations. We then refit the Fe abundance while fixing
[$\alpha$/Fe].

The uncertainty in [Fe/H] includes two components.  The random
component is the 1$\sigma$ error in [Fe/H] from
the $\chi^2$ fitting, accounting for the non-zero covariance between
$T_{\rm eff}$ and [Fe/H].  A systematic error floor of 0.11~dex
is added in quadrature to the random errors for individual stars.
It reflects the non-vanishing difference between DEIMOS
and high-resolution [Fe/H] measurements in the limit of very small
random errors (high SNR). We refer the reader to Kirby et al.\ (2010) and
Vargas et al.\ (2013) for an in-depth description of the analysis.

\begin{figure*}[t]
\plotone{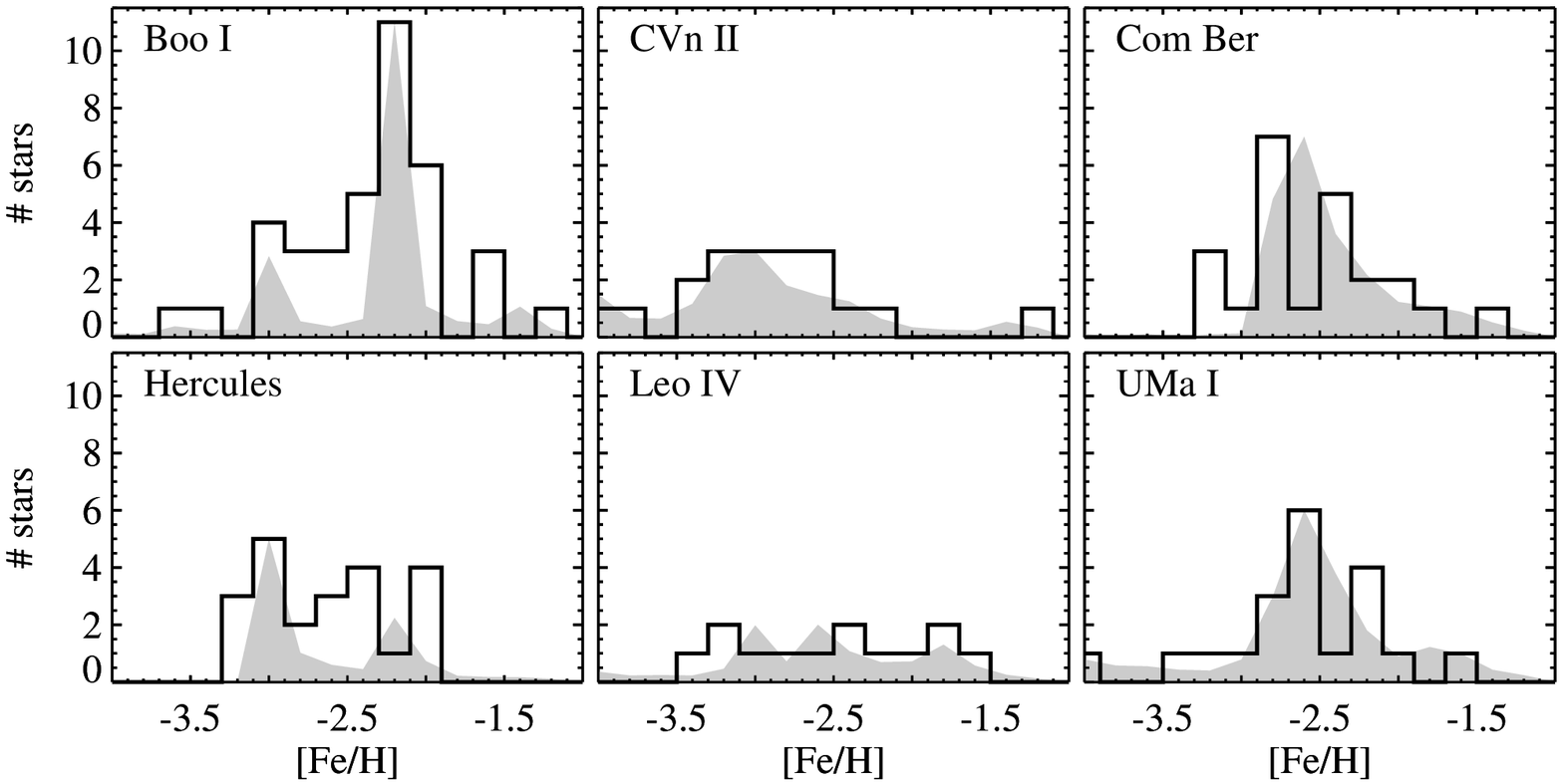}
\caption{
The observed spectroscopic MDF for each UFD (black histograms), along
with an estimate for the PMDF (grey shading; arbitrarily normalized to
the peak of each MDF), given the measured spectroscopic uncertainties.
The observed MDF is used to constrain the SFH fits, while the PMDF is
used to generate Monte Carlo realizations of the MDF in the
characterization of the SFH uncertainties.
The distinctions between the MDF and the PDMF are due to the 
individual metallicity uncertainties for the measurements
comprising each histogram.}
\end{figure*}

\subsubsection{Membership}

We determined the membership status of stars in each sample using
a refined version of the approach adopted
by Simon \& Geha (2007) and Simon et al.\ (2011),
in which all of the available data
for each star, including its velocity, color, magnitude, metallicity,
position, and spectrum, were examined by eye.  Photometry was
extracted from Sloan Digital Sky Survey (SDSS) Data Release 9 
(Ahn et al.\ 2012) or Data Release 10 (Ahn et al.\ 2014) for each
galaxy, and for the particularly sparse UFDs Leo~IV and CVn~II, we
supplemented the SDSS data at faint magnitudes with photometry from
Sand et al.\ (2010) and Sand et al.\ (2012), respectively.  For Boo~I,
Com~Ber, Hercules, and UMa~I, the photometric selection was
based on an $r, g-i$ CMD and an M92 fiducial sequence in similar bands
from Clem (2006).  For Leo~IV, Sand et al.\ (2010) provide much deeper
photometry in $g$ and $r$, so we used an $r, g-r$ CMD and the
corresponding M92 fiducial track.  For CVn~II, the Sand et al.\ (2012)
photometry is in $V$ and $I$, so we transformed the SDSS magnitudes to
those bands with the relations derived by Jordi et al.\ (2006) for
metal-poor stars and compared to a theoretical isochrone (Dotter et al.\
2008) for an age of 12~Gyr and [Fe/H]$ =
-2.21$, which matches the RGB well.  
For Hercules, we also made use of the Str\"omgren photometry
from Ad\'en et al.\ (2009) to separate RGB member stars from foreground
dwarfs.  At bright magnitudes, where the photometric uncertainties are
small ($r \le 20$~mag), the selection window extends 0.1~mag redward and
blueward from the M92 fiducial or isochrone.  At fainter magnitudes,
the SDSS errors increase substantially, and so the selection window is
gradually widened to 0.32~mag away from the fiducial/isochrone at
$r=22.5$~mag.  The Sand et al.\ (2010) photometry for Leo~IV and CVn~II is
deep enough that the photometric uncertainties are negligible even at
the faintest magnitudes of interest for spectroscopy, so the selection
window remains at 0.1~mag at all magnitudes for those galaxies.  Stars
located outside the selection window are considered photometric
non-members, with the exception of one star in Hercules -- a known
spectroscopic member from Koch et al.\ (2008), despite being 0.11~mag redder
than the M92 track.

Stars with velocities more than three standard deviations away from
the galaxy's systemic velocity were classified as non-members, with
the exception of suspected binaries (based
on large velocity differences compared to previous measurements).
We only measure metallicities for two
of these velocity outliers, both RGB stars in Boo~I with velocities
that vary by more than 30~km~s$^{-1}$ from Koposov et al.\ (2011); the
remainder cannot be constrained by our data because of their high
temperatures and/or the low SNR of their spectra.  
We do not make hard cuts on position, metallicity, or \ion{Na}{1}
equivalent width, but stars that are outliers (even if not beyond the
formal limits in color or velocity) in multiple categories are less
likely to be judged as members.  

Our final metallicity distribution function (MDF) for each galaxy was
constructed from the set of RGB stars with valid [Fe/H] fits, relatively low
surface gravities (log~$g < 3.6$), and secure membership.  In these
metal-poor galaxies, HB stars tend to fall far to the blue of the RGB.
Blue HB stars (hotter than $T_{\rm eff} = 11,500$~K) exhibit abundance
anomalies due to atmospheric diffusion (e.g., Grundahl et al.\ 1999),
and are excluded from our sample, but a few red HB stars overlapping
with the RGB may be included.  Membership for a
large majority of the observed stars is obvious and thus secure, but
there will always be stars whose membership is more ambiguous.  For
example, some stars are near the edge of the color selection region,
their velocities are several standard deviations away from the
systemic velocity, and/or they are located at large radii, any of which
increases the likelihood of confusing a foreground star with a UFD
member.  Fortunately, if we include the handful of stars where
membership is questionable, the resulting MDFs are not significantly
changed, and the effect on the SFH fitting is small.  The MDFs for
each galaxy are shown in Figure~3, using a metallicity grid spanning
[Fe/H]~=~$-4.0$ to $-1.0$ with 0.2 dex spacing, matching the
metallicity grid of the isochrone set used for the SFH fits.

\subsubsection{Modeling the Metallicity Distribution Function}

To account for the MDF uncertainties in our SFH fitting, we used a
Bayesian approach to construct a probabilistic MDF (PMDF) associated
with each UFD, where the probabilities are those for the true intrinsic
MDF.  The PMDF enables the generation of artificial MDFs through Monte
Carlo realizations.  We constructed the PMDF as a piecewise constant
function on the same metallicity grid defined for the observed MDFs
and employed in the SFH fits.  The likelihood for the true metallicity
of each star is approximated as a Gaussian that is centered on the
measured metallicity, with a width matching the metallicity error.
The relative weights in the PMDF were estimated using an adaptive
Markov Chain Monte Carlo (MCMC) algorithm, run 10 times for each UFD, with
$10^6$ realizations per run. We then constructed the PMDF from the
draws beyond the first $10^5$ in each run, after the draws had
stabilized.  The resulting PMDFs are shown in Figure~3.  In general,
they match the MDFs well, but there are distinctions because the MCMC
takes into account the distinct measurement errors on individual
stars.  The latter are very heterogeneous, depending upon several
factors (magnitude of the star, observing conditions, metallicity,
etc.).  For this reason, the direct comparison of the MDF and the PDMF
can be slightly deceptive, because the histogram hides the true error
distribution.

\section{Analysis}

\subsection{Comparison with M92}

An inspection of the photometric (Figure~1) and spectroscopic (Figure~3) 
data demonstrates that the stellar populations in our UFD sample
are ancient and metal-poor.  Before we explore the quantitative SFH
fitting for each UFD, it is worth making a comparison to a
well-studied population.  An appropriate object is the Galactic
globular cluster M92 -- one of the most ancient and metal-poor stellar 
systems known.  Of the globular clusters with little extinction, it is 
the most metal-poor (Harris 1996), and it has served as a reference population 
in previous studies of UFDs (e.g., Belokurov et al.\ 2006, 2007, 2009, 2010; 
Okamoto et al.\ 2008, 2012; Sand et al.\ 2009, 2010, 2012).
It was observed with the same camera and filters by Brown et
al.\ (2005), and its CMD is shown in Figure~4.  We assume the cluster
has a true distance modulus of $(m-M)_{\rm o} = 14.62$~mag, taking the
mean of measurements from Paust et al.\ (2007; 14.60$\pm$0.09~mag),
Del Principe et al.\ (2005; 14.62$\pm$0.1~mag), and Sollima et
al.\ (2006; 14.65$\pm$0.1~mag).  We assume $E(B-V)$~=~0.023~mag
(Schlegel et al.\ 1998), and [Fe/H]~=~$-2.3$ (Harris et al.\ 1996).
Comparison of the M92 CMD to those in our UFD sample requires that M92
be shifted in distance and reddening to match those parameters for each UFD.

We determine the distance and reddening to each UFD in our sample by
fitting the HB luminosity and the MS color for stars more than 0.5~mag
below the turnoff (and thus insensitive to age assumptions).  In our
preliminary analysis of 3 galaxies in this sample, we used the RGB
instead of the lower MS to constrain the color (Brown et al.\ 2012),
but the RGB is not well populated in all of our CMDs, and suffers from
significant field contamination, so we altered our approach here.  For
the HB fit, we used the empirical HB locus for M92, because the
metallicity of the cluster falls within the MDF for each galaxy, and
the HB luminosity is a well-known standard candle.  For the MS fit, we
used synthetic MS loci, constructed from 13~Gyr isochrones (VandenBerg
et al.\ 2012), assuming the MDF for each galaxy (Figure~3), a binary
fraction of 48\% (Geha et al.\ 2013), and the photometric errors
determined via the artificial star tests.  Unfortunately, no HB stars
were detected in Com~Ber, and so the fit is only constrained by the
lower MS, resulting in larger uncertainties.  Our derived distances
and reddenings are listed in Table~1.  For Hercules, Leo~IV, and
UMa~I, the values are extremely close to those we determined in our
preliminary analysis of these galaxies (Brown et al.\ 2012), but not
identical, due to the reprocessed photometry and distinct fitting
method here.  The distance and reddening uncertainties are only those
associated with the fits to our UFD photometry, and do not include
systematic errors associated with the M92 distance and reddening, the
isochrones, or MDFs.  For example, the distance to M92 is uncertain at
the level of $\lesssim$0.1~mag, and if we adopted a distinct M92
distance, all of our distances would shift accordingly.  The apparent
distance moduli, $(m-M)_V$, are in good agreement with the values
collected by Martin et al.\ (2008b), although in general our
reddenings are larger and distances are smaller.  More recent
measurements for four of our galaxies (Musella et al.\ 2009, 2012;
Moretti et al.\ 2009; Garofalo et al.\ 2013) also report similar
apparent distance moduli through a combination of larger reddenings
and smaller distances.  If, instead, we were to adopt a combination of
larger distances and smaller reddenings, both M92 and the isochrones
would fall too far to the blue, relative to the MS and RGB in each UFD CMD.

\begin{figure}[t]
\plotone{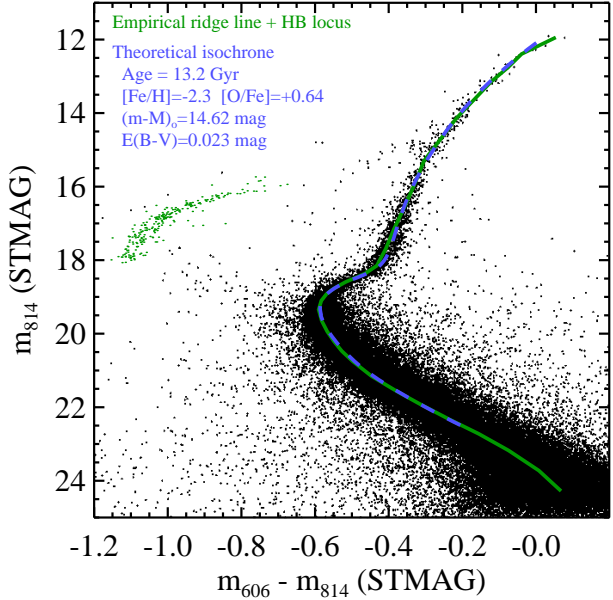}
\caption{The CMD of M92 (Brown et al.\ 2005), observed in the same
  bands on the same camera employed for the UFD observations.  We show
  the empirical ridge line along the MS, SGB, and RGB (green curve),
  along with the HB locus highlighted (green points), which can be
  used as an empirical template for comparison to the UFDs.  We also
  show a theoretical isochrone at the M92 metallicity (blue curve, dashed)
  with excellent agreement for an age of 13.2 Gyr, given the M92
  parameters assumed here (distance, reddening, and composition).  }
\end{figure}

The comparisons between the CMD of M92 and that of each UFD are shown
in Figure~1.  Due to the scarcity of HB stars in each UFD CMD, these
can be shown on top of the HB locus of M92 without confusion.
However, the earlier evolutionary phases in each UFD are well
populated, so for clarity, the MS-SGB-RGB stellar locus of M92 is
shown as a ridge line (see Figure~4; Brown et al.\ 2005).  Although
there are few HB stars in the CMD of each UFD, there is good
agreement between these HB stars and those of M92, because 
the distance to each UFD was determined using the HB as a standard candle.  
Comparing the MS 
turnoff and SGB of M92 to those of each UFD (Figure~1), there is
agreement to first order, implying that the dominant population in
each UFD is as old as the metal-poor globular clusters of the Milky
Way.  However, the UFD stars in the vicinity of the turnoff extend
bluer and brighter than the M92 ridge line, as one would expect from
their MDFs (Figure~3), which extend to metallicities well below that
of M92.  The UFD RGB stars also scatter to the blue of the M92 ridge
line, although it is difficult to quantify, given the contribution of
the asymptotic giant branch stars and field contamination.
Furthermore, the lower MS of each UFD scatters to the red of the M92
ridge line, but this is because of the difference in binary fraction.
Like the dwarf spheroidals (e.g., Minor 2013) and the Galactic field
(e.g., Duquennoy \& Mayor 1991), the UFDs have a binary fraction of
nearly 50\% (Geha et al.\ 2013) -- much higher than the binary
fraction in M92 ($\sim$2\%; Milone et al.\ 2012), which has been
reduced through dynamical evolution (e.g., Ivanova et al.\ 2005).
To explore the UFD CMDs further, we proceed to synthetic CMD analysis.

\subsection{Comparison with Isochrones}

Globular clusters are useful empirical population templates for
comparison to the UFDs, but the known clusters do not span the full
range of age and metallicity required to quantitatively analyze the
UFD populations.  In particular, the UFD populations extend to much
lower metallicities (Figure~3). For this reason, our quantitative
analysis employs theoretical models.  To generate these
models, we use the {\it Victoria-Regina} isochrone and interpolation
codes (VandenBerg et al.\ 2012; VandenBerg et al.\ 2014a), which were
developed for a wide range of stellar population studies, but have a
long history in the study of old metal-poor populations
(e.g., Bergbusch \& VandenBerg 1992; VandenBerg et al.\ 2000;
VandenBerg et al.\ 2006).

We calculated an isochrone grid spanning $-1 > $~[Fe/H]~$ > -4$, with
0.2 dex steps, and $8 < $~age~$< 14.5$~Gyr, with 0.1~Gyr steps.  The
{\it Victoria-Regina} library is available with both scaled-solar
abundances and an enhancement of +0.4 for the $\alpha$-elements (O,
Ne, Mg, Si, S, Ca, and Ti).  We assume [$\alpha$/Fe]~=~+0.4, as
appropriate for old metal-poor populations, such as those in the
Galactic halo and satellites.  While this is certainly appropriate for
most of the UFD population, for the minority of stars at [Fe/H]$> -2$,
there is some indication that the UFDs may have [$\alpha$/Fe] values
that are 0.1--0.2~dex lower (Vargas et al.\ 2013).  If we adopted such
$\alpha$-element abundances for the most metal-rich stars, the ages
for such stars in our fits would be $\sim$0.2--0.4~Gyr older, specifically
due to the change in oxygen abundance, which affects the rate of
the CNO cycle.  Because of its impact on nucleosynthesis (rather than
opacity), the oxygen abundance affects the MS lifetime, and thus the
relation between turnoff luminosity and age.  For the analysis here,
we calculated new grids with the oxygen abundance enhanced beyond the
abundances of the other $\alpha$-elements.  In stars of the diffuse
halo, [O/Fe] appears to increase at decreasing metallicities (Figure~5; 
Frebel 2010).  The isochrone that matches a particular CMD will be
younger as the oxygen abundance increases, with a difference of
$\sim$1~Gyr per 0.5~dex change in [O/Fe] (Figure~6).  The measurements
of [O/Fe] vs.\ [Fe/H] have significant scatter, such that our adopted
[O/Fe] values are uncertain at the level of $\sim$0.2~dex,
corresponding to an absolute age uncertainty of $\sim$0.4~Gyr.
However, if we were to adopt a standard [O/Fe] of +0.4, as frequently
assumed for all $\alpha$-elements when modeling old populations, the
resulting ages in our SFH fits would be significantly older.  Given
their utility in the study of metal-poor populations, the isochrones
with larger [O/Fe] values will be published in a later paper
(VandenBerg et al., in prep.).

\begin{figure}[t]
\plotone{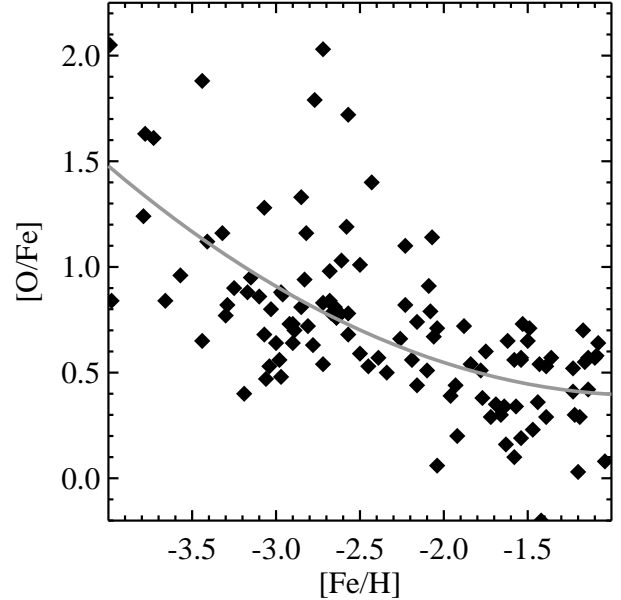}
\caption{The oxygen abundance as a function of metallicity (black points), 
as observed for metal-poor stars in the halo and satellites of the Milky Way 
(Frebel 2010), on the Asplund et al.\ (2009) abundance scale.  
The variation in oxygen abundance adopted in our 
fits comes from a polynomial fit to these data (grey curve).}
\end{figure}

The transformation of the {\it Victoria-Regina} isochrones into the
ACS bands is done via a method similar to that of Brown et
al.\ (2005), although the transformation has been revised to account for
subsequent updates to the isochrone code (VandenBerg et al.\ 2012) and
the library of synthetic spectra employed (Gustafsson et al.\ 2008).
Compared to the previous version of the isochrone code (VandenBerg et
al.\ 2006), the current version includes the effects of He diffusion,
new H-burning nuclear reaction rates, and the adoption of the
Asplund et al.\ (2009) solar metals mixture.  With these updates and
our assumed parameters for M92, the isochrones match the M92 CMD at an
age of 13.2~Gyr (Figure~4), and so the ages in our SFH fits to the UFD CMDs 
should be considered as {\it relative} to this age of M92.  The absolute
age of M92 is itself uncertain at the level of $\sim$1~Gyr, given the
uncertainties in composition, reddening, and distance. For example,
VandenBerg et al. (2014b) prefer a younger age of 12.5~Gyr, due to a
longer assumed distance.  

To fit the observed UFD CMDs, we must convert the isochrone grid into
a set of synthetic CMDs having the same photometric properties as the
observed UFD CMDs.  These photometric properties (scatter and
completeness) were determined via extensive artificial star tests
(\S2.1.3).  Each synthetic CMD was constructed using the {\sc synth}
routine of Harris \& Zaritsky (2001), which takes the isochrone library
and artificial star tests as input.  Each synthetic CMD represents a
stellar population at a single age and metallicity, such that
linear combinations of these synthetic CMDs can be used to fit the
observed UFD CMDs.  The synthetic CMD set for each galaxy is
calculated using the measured distance and reddening values, and
also includes a fixed field contamination component (Table~1).  The
contamination was determined from the Besan\c{c}on Galaxy model (Robin
et al.\ 2003) along the sightline to each galaxy, converted to the ACS
bands using the same synthetic spectra employed in the isochrone
conversion (Gustafsson et al.\ 2008). 

\begin{figure}[t]
\plotone{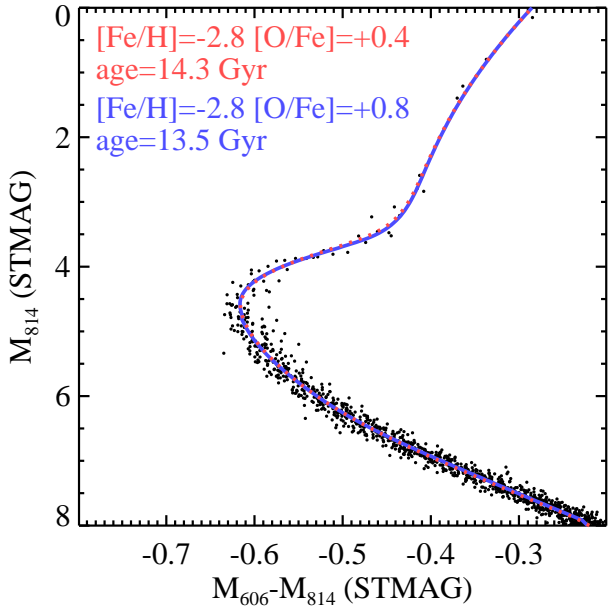}
\caption{A hypothetical CMD for a simple stellar population (black points),
with photometric errors of 0.01~mag in each band.  At a fixed metallicity
([Fe/H]~=~$-2.8$), the CMD can be fit by a younger isochrone (13.5~Gyr;
blue curve) 
with enhanced oxygen abundance ([O/Fe]~=~$+0.8$) or by an older
isochrone (14.3~Gyr; dotted red curve) with the standard
oxygen abundance ([O/Fe]~=~$+0.4$) typically assumed for all
$\alpha$-elements in the fitting of old stellar populations.
}
\end{figure}

\begin{figure*}[t]
\plotone{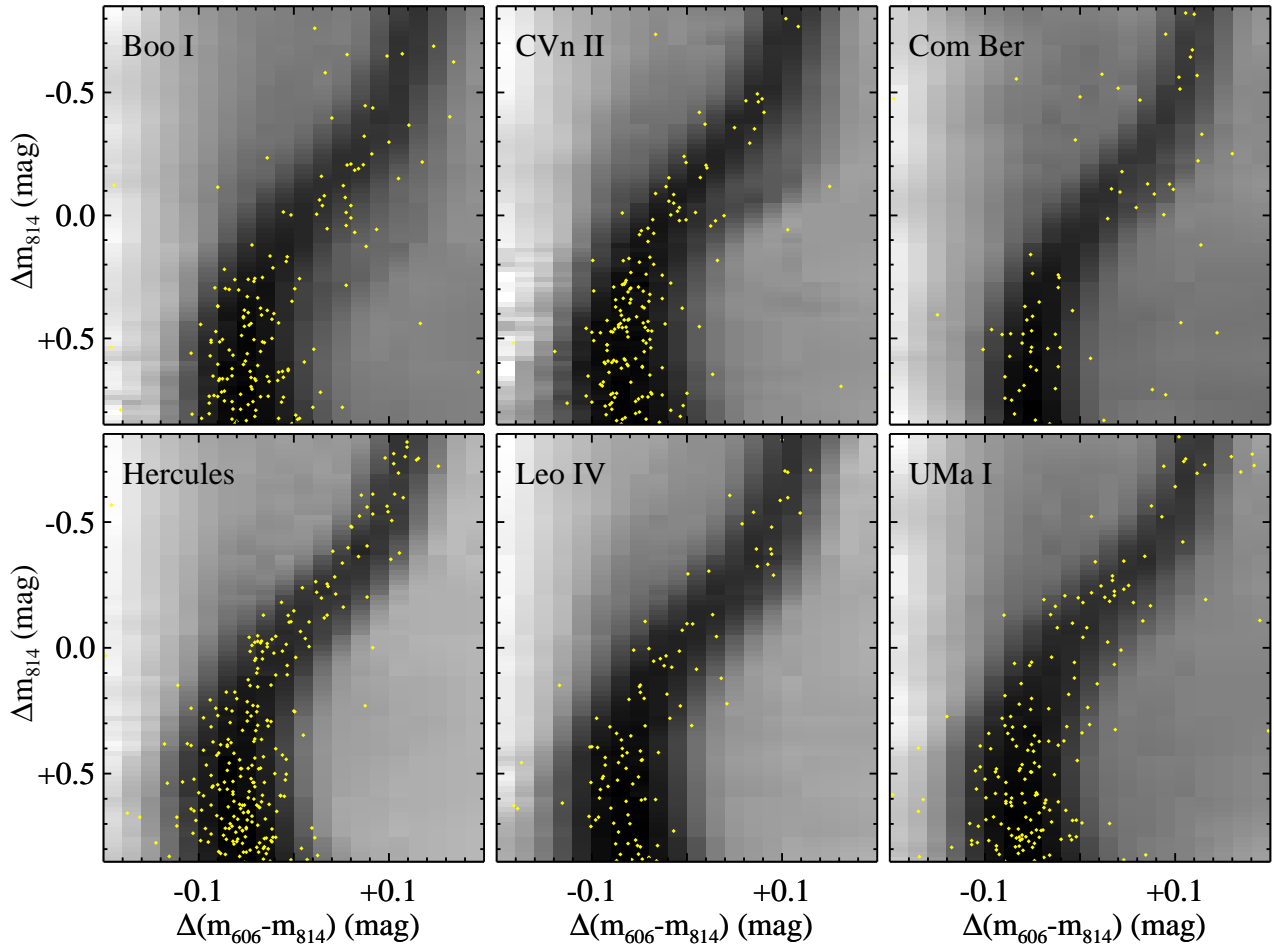}
\caption{
The observed CMD for each UFD (yellow points) compared to the
probability cloud for each associated best-fit model (shading, log
stretch).  The plots here are centered on $M_{606}-M_{814} =
-0.55$~mag and $M_{814} = 3.95$~mag for each galaxy, assuming the
distances and reddenings in Table~1.  The SFH fit is evaluated in a
band 0.2~mag wide that follows the stellar locus and spans
the luminosity range here.  The fit thus concentrates on that part of
the CMD most sensitive to age (the MS turnoff and SGB) while avoiding
those CMD regions that have few or no stars in the data or models.  }
\end{figure*}

We fit the CMDs through the minimization of a Poisson maximum
likelihood statistic (PMLS), and evaluate the best fit through quality
($Q$) and $\chi^2_{\rm eff}$ criteria, each defined in Dolphin (2002;
respectively eqs.\ 10, 23, and 24).  $Q$ evaluates the PMLS of the
best fit (corrected by the number of free parameters) with respect to
the PMLS distribution, and is given in terms of $\sigma$; e.g., $Q =
1$ implies the best fit is 1$\sigma$ worse than a typical fit in the
center of the PMLS distribution.  $\chi^2_{\rm eff}$ is analogous to
the reduced $\chi^2$ value in classical $\chi^2$ minimization, with
values close to unity implying a good fit.  To determine the PMLS
distribution, we perform fits to 10$^4$ Monte Carlo 
realizations of the photometric and spectroscopic data for each
galaxy.  The artificial realization of the photometric data is a random draw on
the best-fit CMD model that results from the synthetic CMD fitting of
the observed UFD CMD.  The artificial realization of the spectroscopic data is a
random draw on the PMDF estimated in \S2.2.5.

We restrict our SFH fits to that part of the CMD from the MS turnoff
through the top of the SGB (Figure~7).  By doing so, we avoid those
parts of the CMD insensitive to age, such as the RGB and lower MS,
which would otherwise dilute the impact of age variations on the fit.
The restriction also has other motivations.  By avoiding the lower MS,
we restrict the fit to a small mass range ($\Delta M
<$0.1~$M_\odot$), thus minimizing the sensitivity to the assumed IMF.
By avoiding CMD regions where few or no stars are observed, and where
few or no stars are predicted by the models, we prevent artificially
enhancing the quality of the best fit, because the agreement between data
and models in empty CMD regions is irrelevant (e.g., see
discussion in Dolphin 2002).  Finally, we avoid the BS sequence, which
would otherwise mimic a minority population component far younger than
the dominant population; as explained in \S2.1.4, there are a few BS
stars in each UFD CMD, with the ratio of BS to HB stars $\sim$2,
similar to that observed in the Galactic halo and other dwarf
galaxies.  The identical region is fit in the CMD of each galaxy, with
the region boundaries shifted from galaxy to galaxy using the
reddening and distance to each galaxy.  For the IMF power-law slope
and binary fraction, we assume $dN / dm \propto m^{-1.2}$ and 48\%,
respectively, previously derived for Hercules (Geha et al.\ 2013).
However, because the fitting region is restricted to the upper MS and
SGB, these choices are not important.  For example, values of $-2.2$
for the IMF power-law slope or 38\% for the binary fraction lead to negligible
differences in the resulting SFH ($<$0.2~Gyr shifts in age).

Comparison of the observed CMDs to the synthetic CMDs demonstrates
that the observed CMDs can be reproduced with a very simple model,
comprised of two episodes of star formation.  Specifically, the fit
has 3 parameters: the ages of the two components, and the fraction of
star formation in each.  Each episode is a single-age population, but
can have a range of metallicities.  The metallicities in the fit are fixed to
match the observed spectroscopic MDF, with the constraint that the
metallicity monotonically increases at younger ages.  The parameters
of the best-fit model are listed in Table~2.  In Figure~7, we 
show a comparison of the observed UFD CMDs to the best-fit synthetic CMDs, 
each represented as a two-dimensional distribution of probability density.

In general, the fits are excellent, particularly when one considers
the simplicity of the 3-parameter model and the fact that the
metallicities are constrained in the fits.  Adding two additional
parameters to the fit for each galaxy, varying the duration of star
formation in each of the two bursts, does not improve the fit quality.
The resulting 5-parameter fits minimize the duration of star formation
in each burst and do not improve the PMLS, underscoring the preference
for a narrow age range in each burst.  

The two-burst model is a better match to each CMD than a model with a
single burst, and has the advantage of quantifying the possible
contribution of a minority population.  However, a single burst of
star formation cannot be ruled out from these data.  This is not
surprising, if one inspects the results of Table~2.  For Boo~I, the
two components are essentially the same age.  For CVn~II, Com~Ber,
Hercules, and Leo~IV, the younger component is small ($<$25\% of the
population).  If the SFH fit to each CMD is forced to a single age,
the result is within 0.2~Gyr of the mean age in a two-burst model
(within the uncertainties on the mean age; see Table~2), with
less than 1$\sigma$ of degradation in fit quality.  Furthermore, in
the fits to the Monte Carlo realizations, a small but non-negligible
fraction of the fits ($<$20\% of the time) result in an essentially
single-age population, with both components having ages within 0.5~Gyr
of each other.

\begin{table*}[t]
\begin{center}
\caption{SFH Fitting}
\begin{tabular}{lcrcrccc}
\tableline
                  & Age\tablenotemark{a}   & Fraction & Age\tablenotemark{a}   & Fraction & Mean        &             &  \\
                  & Component 1     & Component 1        & Component 2     & Component 2        & Age\tablenotemark{b}         &     & $Q$          \\
Name              & (Gyr) & (\%)     & (Gyr) & (\%)     & (Gyr)       & $\chi_{\rm eff}$ & ($\sigma$) \\
\tableline
Bootes I          & 13.4  &  3       & 13.3  & 97       & 13.3$\pm$0.3 & 1.05 & +0.9\\ 
Canes Venatici II & 13.8  & 95       & 10.6  &  5       & 13.6$\pm$0.3 & 0.99 & -0.2\\
Coma Berenices    & 14.0  & 96       & 11.1  &  4       & 13.9$\pm$0.3 & 1.09 & +1.8\\
Hercules          & 13.7  & 82       & 10.6  & 18       & 13.1$\pm$0.3 & 0.98 & -0.3\\
Leo IV            & 13.7  & 77       & 11.2  & 23       & 13.1$\pm$0.4 & 1.01 & +0.2\\
Ursa Major I      & 14.1  & 45       & 11.6  & 55       & 12.7$\pm$0.3 & 1.02 & +0.3\\
\tableline
\multicolumn{8}{l}{$^{\rm a}$Relative to an M92 age of 13.2~Gyr.} \\
\multicolumn{8}{l}{$^{\rm b}$Mean age of the two-component model, with statistical uncertainties only.} \\
\end{tabular}
\end{center}
\end{table*}

When comparing the best-fit models for each galaxy (Table~2), the fit
quality is a bit better than expectations for CVn~II and Hercules, but
their PMLS scores are still well within the distribution from the
Monte Carlo runs.  The worst fit is that for Com~Ber, which is
1.8$\sigma$ worse than the median PMLS score in the Monte Carlo runs,
although a 1.8$\sigma$ outlier is not unreasonable for a sample of six
galaxies.  The Com~Ber dataset is by far the most problematic in our
survey.  Despite the large number of tiles used to observe the galaxy,
its CMD is poorly populated, its distance is not well constrained
(with no HB stars in the CMD), and the large number of tiles led to a
high field contamination (24\%).

The uncertainties in the fit can be derived from the Monte Carlo fits
to artificial realizations of the CMD and MDF.  Using the results of
these Monte Carlo fits, the statistical uncertainty on the mean age of
the population is well-defined, and included in Table~2 for each
best-fit model.  However, the uncertainties on the age and fraction
for each of the two population components are not well-defined,
because the fraction and age are strongly correlated.  The older
component has a standard deviation of 0.2--0.6~Gyr in the Monte Carlo
runs for each galaxy.  The age of the younger component varies much
more widely (standard deviations of 1.1--1.8~Gyr), because in many of
the Monte Carlo runs, the younger component is only a trace population
($<$10\%).  For example, in the best-fit model for Boo~I, the two
components are nearly identical in age, with most of the weight in the
slightly younger component (see Table~2).  If we restrict the analysis
to those Monte Carlo runs where this younger component is dominant
($>$50\%), the standard deviation in the age of the second component
is 0.4~Gyr, but if we include those runs where the second component is
only a trace population, the standard deviation is 1.7~Gyr.
For these reasons, the uncertainties on the individual components
are best expressed in a plot of cumulative SFH for each
galaxy, shown in Figure~8.  In such a plot, the fraction of the population
that can fall in the second component quickly dwindles as the age
of this component falls below 12~Gyr.

For each of the best-fit models, a significant fraction of the
population is approximately as old as the universe, as measured in the
9-year results from the {\it Wilkinson Microwave Anisotropy Probe}
({\it WMAP}; 13.75$\pm$0.085~Gyr; Hinshaw et al.\ 2013).  
Although the oldest stars in the best-fit model formally exceed the
age of the universe for Com~Ber and UMa~I, the exceedances are not
significant when one considers the statistical and systematic
uncertainties involved.  As far as the statistical uncertainties are
concerned, there is almost no difference in fit quality between the
models derived above and ones that are bounded by the age of the
universe.  The systematic uncertainties
associated with our modeling are even larger than the statistical
uncertainties, and are primarily related to the oxygen abundance and
distance moduli assumed in the fits.  If we were to assume distance
moduli that are 0.05~mag shorter or longer, the resulting ages would
shift $\sim$0.5~Gyr older or younger, respectively.  If we were to
assume [O/Fe] values that are 0.2~dex lower or higher, the resulting
ages would shift $\sim$0.4~Gyr older or younger, respectively.  For
this reason, the SFH fits we present here are best considered as {\it
  relative} ages with respect to an M92 that is 13.2~Gyr old.

The three most distant galaxies in our sample were also observed with
the Wide Field Planetary Camera 2 (WFPC2) on {\it HST}.  Although the
WFPC2 data are noisier at the MS turnoff than the ACS data we present
here, Weisz et al.\ (2014b) fit the SFHs for these three galaxies from
the WFPC2 CMDs, and found similar results to our own for Hercules and
Leo~IV, with 70\% of the SFH occurring by $\sim$12~Gyr.  In contrast,
they found CVn~II to be significantly younger, with 70\% of the
stars older than $\sim$10~Gyr, and a tail to even younger ages.  The
distinction is puzzling, because WFPC2 observations of CVn~II fall
completely within the ACS observations, albeit with half the areal
coverage, $\sim$4 times less throughput, and $\sim$4 times less
exposure time.  Their finding of younger stars in CVn~II does not seem
to be due to distance assumptions.  Weisz et al.\ (2014a) assume
apparent distance moduli for CVn~II, Hercules, and Leo~IV that are
0.06, 0.05, and 0.2~mag shorter than our own (and similarly shorter
than those of Martin et al.\ 2008b).  All else being equal, this would
make their ages about 0.5~Gyr older than our ages for CVn~II and
Hercules, and about 2~Gyr older for Leo~IV, but the offsets in
distance modulus for Hercules and CVn~II are nearly identical.  They
used a distinct set of isochrones, but this would not give an offset
with only one galaxy.  They assume the color excess from reddening is
about 0.05~mag larger in Hercules than in CVn~II, as do we, so it
cannot be due to a relative color shift.  We assume CVn~II is somewhat
more metal poor than Hercules, and that is the reason we actually find
CVn~II to be 0.5~Gyr older than Hercules (on average), despite the
fact that the ACS CMDs are very similar.  Although they make no
mention of BS stars, there are only a few in the CVn~II CMD, so their
presence would not yield a significantly young population in their
CVn~II fit, even if they were modeled as young stars.  A possible
explanation is the depth of their data.  In the WFPC2 data, the MS
turnoff is closer to the faint limit, and so there is significantly
more spread at the turnoff due to photometric errors, which might
allow a younger population in their fits.  If we appropriately
increase the photometric errors in both the ACS
catalog for CVn~II and its artificial star tests, a wider range of SFHs are
consistent with the data.

\begin{figure*}[t]
\plotone{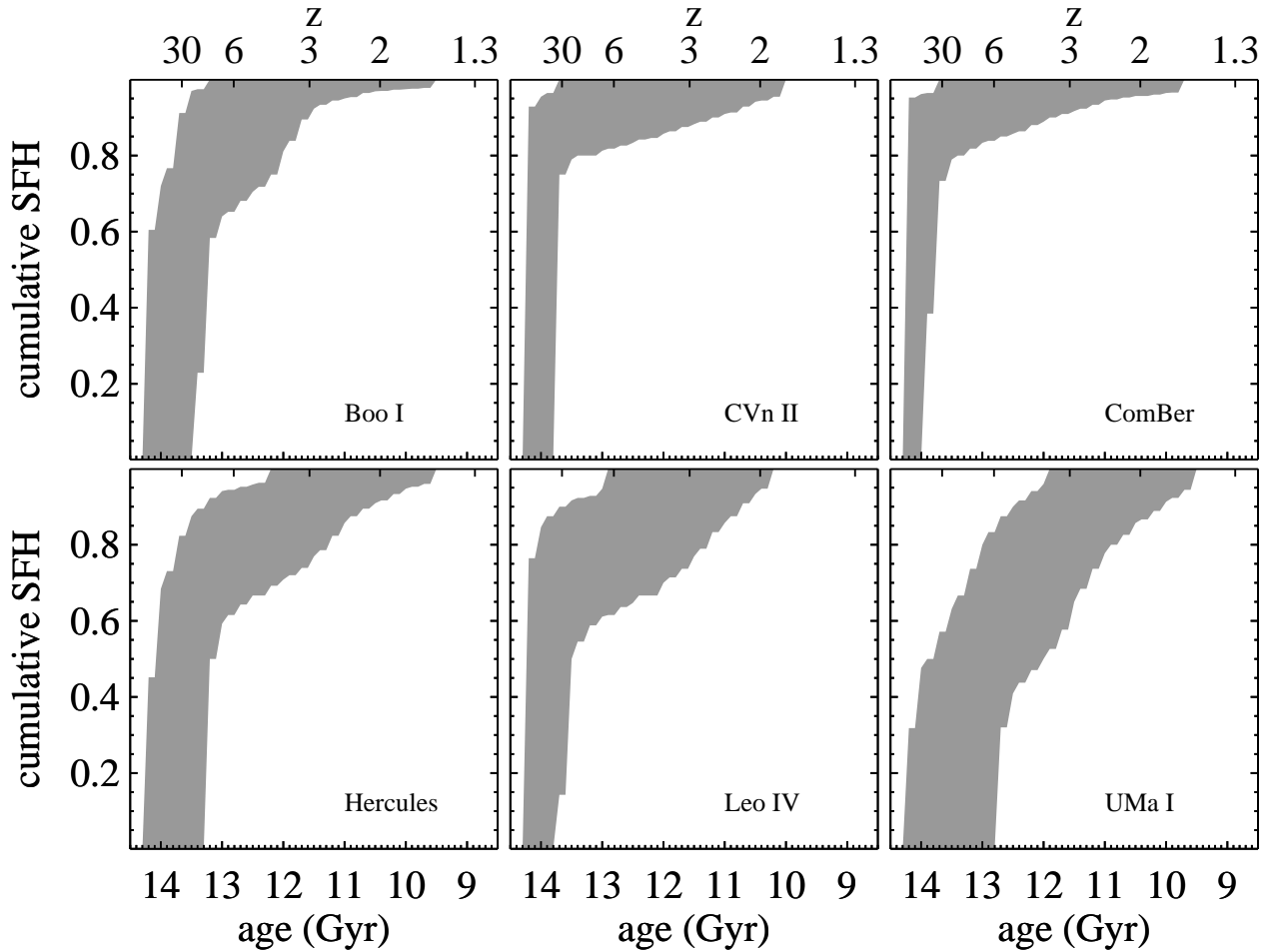}
\caption{The statistical uncertainties for the cumulative SFH for each
  galaxy, assuming two bursts, and determined by 3-parameter fits to
  10$^4$ Monte Carlo realizations of the photometric and spectroscopic
  data.  Within these 1$\sigma$ uncertainties, the SFH for each galaxy
  is consistent with a model that has at least 80\% of the star
  formation completing by $z \sim 6$.  }
\end{figure*}

\section{Discussion}

We have used a combination of {\it Keck} spectroscopy and {\it HST}
photometry to characterize the stellar populations of six faint Milky
Way satellites.  The spectroscopy demonstrates that these galaxies are
comprised of extremely metal-poor stars; the lowest metallicities are
consistent with pre-enrichment from a single supernova (see Wise et
al.\ 2012).  Using these metallicities as a constraint in fits to
high-precision CMDs, we find that each of these galaxies is
well-matched by a population of ancient stars, with no indication of 
a delayed onset to star formation (cf.\ Noeske et al.\ 2007).  
In the best-fit models
for 5 galaxies (Table~2), a majority ($>$75\%) of the stars formed
prior to $z \sim 10$ (13.3~Gyr ago), when the epoch of reionization
began (Hinshaw et al.\ 2013).  Within the uncertainties (Figure~8),
all 6 of the galaxies formed at least 80\% of their stars by $z \sim
6$, although this fraction might be as low as 40\% for UMa~I and 60\%
for Hercules and Leo~IV.  
A two-burst model reproduces the data well, but we cannot
rule out a single ancient burst of star formation in each galaxy.  We
stress that the absolute age scale is uncertain at the level of
$\sim$1~Gyr, given the systematics associated with distance and
abundances.  For example, if we were to adopt a longer distance for
M92, the age of M92 and the UFDs would all shift younger, and if we
were to adopt lower [O/Fe] values, the ages would shift older.  In the
coming decade, more accurate distances from the {\it Gaia} mission
will reduce such systematic errors considerably (Perryman et
al.\ 2001).  In the next few years, we should also have an accurate
{\it HST} parallax to a metal-poor globular cluster (NGC~6397; program
GO-13817), which can then replace M92 as a population template for
this kind of work.

The populations of these galaxies are very similar to each other
(Figures~1 and 2), as one might expect if they were all influenced by
an event that synchronized the truncation of star formation in each.
These faint satellites stand in contrast to the brighter dwarf
spheroidals, all of which host stars younger than 10~Gyr (Orban et
al.\ 2008).  It is worth noting that the UFD SFHs may be even more
abrupt and synchronized than we report here.  Although our relative
ages are robust, the distance uncertainties for each galaxy, and the
scatter in [O/Fe] (both galaxy to galaxy and within a given galaxy)
may manifest as an age spread in our fits.  

The discovery of additional faint satellites around the Milky Way and
Andromeda have narrowed the gap between observations and $\Lambda$CDM
predictions of substructure.  To close that gap, simulations of galaxy
formation assume that reionization suppressed the star formation in
the smallest DM sub-halos (e.g., Bullock et al.\ 2001; Ricotti \& Gnedin
2005; Mu$\tilde{\rm n}$oz et al.\ 2009; Bovill \& Ricotti 2009,
2011a, 2011b; Tumlinson 2010; Koposov et al.\ 2009; Li et al.\ 2010;
Salvadori \& Ferrara 2009; Salvadori et al.\ 2014).  Specifically,
such models assume that reionization heated the gas in small DM halos
to $\sim$10$^4$~K, and the resulting thermal pressure boiled the gas
out of the halos and into the intergalactic medium (IGM).  Gravity is
too weak in these sub-halos to retain the gas or reacquire it from the
reionized IGM.  The stellar populations of the UFDs, which are
extremely similar to each other 
and dominated by ancient metal-poor stars, support
the premise of an early synchronizing event in their SFHs.
Although galaxy formation 
models tune the suppression threshold in terms of DM mass,
the outcome is manifested in terms of luminous matter, with
post-reionization star formation plummeting in satellites fainter than
$M_V \sim -8$~mag.  Outside of simulations, the threshold is likely
not as clean as this, with multiple parameters affecting the outcome,
including the details of the star formation history, the DM accretion
history, local dynamics, metallicity, location within the parent halo,
and distance from major sources of reionization.  It is difficult to
disentangle such effects with the small sample here.  For example,
Boo~I and Com~Ber have almost exclusively old populations, and fell
into the Milky Way earlier than the other galaxies in our sample
(Rocha et al.\ 2012), giving them an earlier exposure to the dominant
source of ionization.  While UMa~I is dominated by old metal-poor
stars, it appears to be systematically younger than the other galaxies
in our sample.  UMa~I may be distorted, and Okamoto et al.\ (2008) argue
that it appears to be undergoing disruption; elongation along our
sightline could be broadening the CMD, producing an apparent age
spread.  Hercules is the brightest galaxy in our sample ($M_v =
-6.6$~mag; Martin et al.\ 2008b); it may have retained more gas during
the reionization era, leading to a non-negligible population of
younger stars.  Besides the galaxies in our sample, 
there are others that demonstrate these complexities.
For example, Leo~T is a gas-rich irregular hosting recent star formation,
despite having a luminosity similar to those of the ancient UFDs
(Irwin et al.\ 2007; de Jong et al.\ 2008a; Ryan-Weber et al.\ 2008);
at 409~kpc, its isolation from the Milky Way could have enabled its
evolution as a ``rejuvenated fossil,'' with late gas accretion and
associated star formation (Ricotti 2009).

With the current facilities, measuring SFHs with
cosmologically-interesting constraints can only be done for stellar
populations within the Local Group.  Unfortunately, we only know of a
few Milky Way satellites near $M_V \sim -8$~mag, where we might better
understand the conditions that lead to a reionization-induced
suppression of star formation.  Increasingly faint dwarfs are also being
discovered at $z \sim 1-2$ (e.g., Atek et al.\ 2014; Alavi et al.\ 2014),
but these have stellar masses that are several orders of magnitude larger
than the UFD satellites of the Milky Way. Because these intermediate-redshift
galaxies are well above the filtering mass, they should not experience the 
quenching effects of reionization, and in fact exhibit significant star 
formation beyond $z \sim 6$.  In the near future, the best hope for
further progress in this area comes from additional wide-field surveys
that should reveal additional faint satellites (Willman 2010), such as
the {\it Panoramic Survey Telescope And Rapid Response System}, 
the {\it Dark Energy Survey}, the {\it Large Synoptic Survey Telescope}, 
and the {\it Wide Field Infrared Survey Telescope}.  Satellites found
in these surveys would be prime targets for both {\it HST} and the {\it
  James Webb Space Telescope}.

\acknowledgements

Support for program GO-12549 was provided by NASA through a grant from
the Space Telescope Science Institute, which is operated by the
Association of Universities for Research in Astronomy, Inc., under
NASA contract NAS 5-26555.  This work was supported by a NASA Keck PI
Data Award, administered by the NASA Exoplanet Science Institute under
RSA number 1474359.  Data presented herein were obtained at the
W.M. Keck Observatory from telescope time allocated to NASA through
the agency's scientific partnership with the California Institute of
Technology and the University of California. The Observatory was made
possible by the generous financial support of the W.M. Keck
Foundation.  The authors wish to recognize and acknowledge the very
significant cultural role and reverence that the summit of Mauna Kea
has always had within the indigenous Hawaiian community. We are most
fortunate to have the opportunity to conduct observations from this
mountain.  We thank the anonymous referee who suggested revisions
that improved the clarity of this work.
We are grateful to P.\ Stetson for providing his DAOPHOT-II
code and offering assistance with its use.  D.A.V acknowledges the
support of a Discovery Grant from the Natural Sciences and Engineering
Research Council of Canada.  L.C.V. was supported by the National
Science Foundation Graduate Research Fellowship under Grant
No. \mbox{DGE$-$1122492}.  R.R.M. acknowledges partial support from
CONICYT Anillo project ACT-1122 and project BASAL PFB-$06$, as well as
FONDECYT project N$^{\circ}1120013$.  P.G. acknowledges support from
NSF grant AST-1010039.

\end{document}